\documentclass[structabstract]{aa}  
\usepackage{graphicx}
\usepackage{txfonts}
\begin{document}
   \title{Thermo-physical properties of 162173 (1999~JU3), a potential flyby
          and rendezvous target for interplanetary missions}
%   \subtitle{Based on the experience from the Hayabusa flyby target 25143~Itokawa}

   \author{T.\ G.\ M\"uller\inst{1}
           \and
           J.\ \v{D}urech\inst{2}
           \and
           S.\ Hasegawa\inst{3}
           \and
           M.\ Abe\inst{3}
           \and
           K.\ Kawakami\inst{3}
           \and
           T.\ Kasuga\inst{4}
           \and
           D.\ Kinoshita\inst{5}
           \and
           D.\ Kuroda\inst{6}
           \and
           S.\ Urakawa\inst{7}
           \and
           S.\ Okumura\inst{7}
           \and
           Y.\ Sarugaku\inst{8}
           \and
           S.\ Miyasaka\inst{9}
           \and
           Y.\ Takagi\inst{10}
           \and
           P.\ R.\ Weissman\inst{11}
           \and
           Y.-J.\ Choi\inst{12}
           \and
           S.\ Larson\inst{13}
           \and
           K.\ Yanagisawa\inst{6}
           \and
           S.\ Nagayama\inst{14}
          }

\institute{%
    Max-Planck-Institut f\"ur extraterrestrische Physik,             %1
    Giessenbachstra{\ss}e, 85748 Garching, Germany;
    \email{tmueller@mpe.mpg.de}
 \and
    Astronomical Institute, Faculty of Mathematics and Physics,      %2
    Charles University in Prague, 
    V Hole\v{s}ovi\v{c}k\'{a}ch 2, 180 00, Praha 8, Czech Republic;
 \and
    Institute of Space and Astronautical Science,                    %3
    Japan Aerospace Exploration Agency,
    3-1-1 Yoshinodai, Sagamihara, Kanagawa 229-8510, Japan;
 \and
    Department of Physics and Astronomy, The University of Western       %4
    Ontario, 1151 Richimond St. London, ON, N6A 3K7, Canada;
 \and
    Institute of  Astronomy, National Central University,            %5
    300 Jhongda RD., Jhongli, Taoyuan 32001, Taiwan;
 \and
    Okayama Astrophysical Observatory, National Astronomical         %6
    Observatory of Japan, 3037-5 Honjo, Kamogata, Asaguchi,
    Okayama 719-0232, Japan;
 \and
    Bisei Spaceguard Center, Japan Spaceguard Association,         %7
    1716-3 Okura, Bisei, Ibara, Okayama 714-1411, Japan;
 \and
    Kiso Observatory, Institute of Astronomy,                        %8
    The University of Tokyo, 10762-30 Mitake, Kiso,
    Nagano 397-0101, Japan;
 \and
    Tokyo Metropolitan Government, 2-8-1 Nishishinjyuku,             %9
    Shinjyuku, Tokyo 163-8001, Japan
 \and
    Aichi Toho University, 3-11 Heiwagaoka, Meito,           %10
    Nagoya, Aichi 468-8515, Japan;
 \and
    Jet Propulsion Laboratory, 4800 Oak Grove Drive,                %11
    MS 183-301, Pasadena, CA 91109, USA;
 \and
    Korea Astronomy and Space Science Institute,                     %12
    61-1 Hwaam-dong, Yueing-gu, Daejeon, Taejon 305-348, Korea;
 \and
    Lunar and Planetary Laboratory, University of Arizona,          %13
    Tucson, AZ, 85721-0092, USA;
 \and
    National Astronomical Observatory of Japan,                     %14
    2-21-1 Osawa, Mitaka, Tokyo 181-8588, Japan;
    }

   \date{Received; accepted}

% \abstract{}{}{}{}{} 
% 5 {} token are mandatory
 
  \abstract
   {Near-Earth asteroid 162173 (1999~JU3) is a potential flyby and rendezvous target
    for interplanetary missions because of its easy to reach orbit. The physical
    and thermal properties of the asteroid are relevant for establishing the
    scientific mission goals and also important in the context of near-Earth
    object studies in general.}
   {Our goal was to derive key physical parameters such as shape, spin-vector,
    size, geometric albedo, and surface properties of 162173 (1999~JU3).}
   {With three sets of published thermal observations (ground-based N-band, Akari IRC,
    Spitzer IRS), we applied a thermophysical model to derive the radiometric properties
    of the asteroid.
    The calculations were performed for the full range of possible shape and spin-vector
    solutions derived from the available sample of visual lightcurve observations.}
   {The near-Earth asteroid 162173 (1999~JU3) has an effective diameter of
    0.87\,$\pm$\,0.03\,km and a geometric albedo of 0.070\,$\pm$\,0.006.
    The $\chi^2$-test reveals a strong preference for a retrograde sense of rotation
    with a spin-axis orientation of $\lambda_{\mathrm{ecl}}\,=\,73^{\circ}$, $\beta_{\mathrm{ecl}}\,=\,-62^{\circ}$ and P$_{\mathrm{sid}}$\,=\,7.63$\pm$0.01\,h.
    The most likely thermal inertia ranges between 200 and
    600\,J\,m$^{-2}$\,s$^{-0.5}$\,K$^{-1}$, about a factor of 2 lower than the value
    for 25143~Itokawa. This indicates that the surface lies somewhere between a thick-dust regolith
    and a rock/boulder/cm-sized, gravel-dominated surface like that of 25143~Itokawa.
    Our analysis represents the first time that
    shape and spin-vector information has been derived from a combined data set of 
    visual lightcurves (reflected light) and mid-infrared photometry and spectroscopy
    (thermal emission).}
   {}


  \authorrunning{T.\ G.\ M\"uller et al.}
  \titlerunning{Thermo-physical properties of 162173 (1999~JU3)}
   \maketitle
%
%________________________________________________________________

\section{Introduction}

Asteroid \object{162173 (1999~JU3)} is currently among the potential targets
of future interplanetary exploration missions. The target is relatively easy
to reach with state-of-the-art mission capabilities, and it offers high
scientific potential (Binzel et al.\ \cite{binzel04}).
This near-Earth asteroid belongs to the
C-class objects, which are believed to represent primitive, volatile-rich
remnants of the early solar system. 
Various aspects of this small body have been covered in some detail in
the recent works by Hasegawa et al.\ (\cite{hasegawa08}) and by
Campins et al.\ (\cite{campins09}). 

Hasegawa et al.\ (\cite{hasegawa08}) use a spherical shape model for
their radiometric analysis, and alternatively use an ellipsoidal shape model,
but without knowing the true spin-vector orientation. The results
(radiometric diameter of 0.92\,$\pm$\,0.12\,km, visual geometric albedo of
0.063$^{+0.020}_{-0.015}$) which indicate a thermal inertia larger than
500\,J\,m$^{-2}$\,s$^{-0.5}$\,K$^{-1}$),
were based on a set of photometric Subaru and {\it Akari} observations and 
connected to simplified shape and spin-axis assumptions.
Campins et al.\ (\cite{campins09}) have obtained a single-epoch {\it Spitzer}
infrared spectrograph (IRS) spectrum. They used a spherical shape model, and for the spin-pole
orientation they used the extreme case of an equatorial retrograde
geometry and a prograde solution published by Abe et al.\ (\cite{abe08}). Their analysis, based on the
single IRS-spectrum and ignoring the data sets published by
Hasegawa et al.\ (\cite{hasegawa08}), yielded a value for the thermal inertia of
700\,$\pm$\,200\,J\,m$^{-2}$\,s$^{-0.5}$\,K$^{-1}$,
a diameter estimate of 0.90\,$\pm$\,0.14\,km, and geometric albedo of 0.07\,$\pm$\,0.01.

Despite the simplifications in shape and spin-vector properties, both
sets of published radiometric diameter and albedo values agree within
the given uncertainties. Both teams also favoured a relatively
high thermal inertia (close to that of \object{25143~Itokawa}).
The unknowns of the spin-vector orientation cause
a large uncertainty in the thermal properties.

Here we re-analyse all available lightcurve observations to derive 
(on the basis of standard $\chi^2$ lightcurve inversion techniques) all
matching spin-vector and shape solutions (Sect.~\ref{sec:sv}). The full
possible range for shape, spin-axis orientation, and rotation period was
then used as input for a thermophysical $\chi^2$ analysis
of all available thermal observations (Sect.~\ref{sec:tpm}) with the goal
of deriving radiometric sizes, albedos and thermal inertias. At the same
time, we determined the most likely shape-solution, rotation period
and spin-axis orientation (Sect.~\ref{sec:results}).

%__________________________________________________________________

\section{Possible shape and spin-vector solutions}
\label{sec:sv}

\begin{figure}[h!tb]
    \begin{center}
      \resizebox{\hsize}{!}{\includegraphics{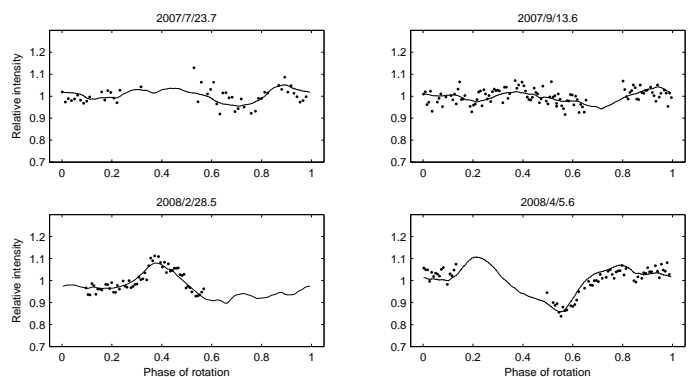}}
      \caption{Match between observed lightcurves and shape/spin-axis model
       solution ``7\_1".
       \label{fig:lc}}
    \end{center}
  \end{figure}

   A detailed list of the available photometric observations is presented
   in Table~\ref{tbl:lc_obs}. There are about 40 dedicated visual
   lightcurve data sets spread over more than 270 days. About half
   of the lightcurves were calibrated; some of the data sets
   are very noisy. Based on these data,
   Abe et al.\ (\cite{abe08}) found a rotation period of
   7.6272\,$\pm$\,0.0072\,h
   and a spin orientation of $\lambda_{\mathrm{ecl}}\,=\,331.0^{\circ}$, $\beta_{\mathrm{ecl}}\,=\,+20.0^{\circ}$,
   indicating a prograde rotation. A more recent analysis by the same authors (priv.\
   communication) resulted in a rotation model with slightly different values:
   $\lambda_{\mathrm{ecl}}\,=\,327.3^{\circ}$, $\beta_{\mathrm{ecl}}\,=\,+34.7^{\circ}$, P$_{\mathrm{sid}}$\,=\,7.6273922\,hours.
   Both solutions were derived using the epoch and amplitude methods
   described by Magnusson (\cite{magnusson86}). These methods are reliable for
   irregularly shaped bodies and for sufficient lightcurve data covering various
   aspect angles. \object{1999~JU3} has a comparatively spherical shape
   and the available lightcurve data were from restricted directions.
   Nevertheless, both solutions gave a reasonable match to the observed lightcurves,
   but it turned out that these solutions are not unique and other parameter sets
   with different rotation periods, spin-axis orientations and shape models
   could not be excluded (see Fig.~\ref{fig:lc}).
   
  \begin{table}[h!tb]
    \begin{center}
    \caption{Observation circumstances for the lightcurve measurements
             (see also Table.~1 in Abe et al.\ \cite{abe08}).
             \label{tbl:lc_obs}}
    \begin{tabular}{lll}
      \hline
      \hline
      \noalign{\smallskip}
      Mon/Day (2007)  &  Telescope & Observer \\
      \noalign{\smallskip}
      \hline
      \noalign{\smallskip}
      07/8, 09/4                 & 2.2\,m/Mauna Kea   & T.\ Kasuga \\
      \noalign{\smallskip}
      07/19-23, 12/3,4,6-8,      & 1.0\,m/Lulin       & M.\ Abe, K.\ Kawakami, \\
      02/26-28, 04/2,4,5         &                    & D.\ Kinoshita \\
      \noalign{\smallskip}
      08/5,15, 09/6,11,13,15,    & 1.0\,m/Ishigaki    & D.\ Kuroda, S.\ Nagayama, \\
      10/6,18, 11/13,15          &                    & K.\ Yanagisawa \\
      \noalign{\smallskip}
      08/9-10,17,20, 09/6,10     & 1.0\,m/Bisei       & S.\ Urakawa, \\
                                 &                    & S.\ Okumura \\
      \noalign{\smallskip}
      09/4,5,7,8,10,12,14,15     & 1.05\,m/Kiso       & M.\ Abe, K.\ Kawakami, \\
      11/7-9,11,13, 02/5-8,      &                    & Y.\ Sarugaku, Y.\ Takagi, \\
      04/14,15                   &                    & S.\ Miyasaka \\
      \noalign{\smallskip}
      09/11-14                   & 1.55\,m/Steward    & P.\ R.\ Weissman, \\
                                 &                    & Y.-J.\ Choi, S.\ Larson \\
     \noalign{\smallskip}
     \hline
     \noalign{\smallskip}
    \end{tabular}
    \end{center}
  \end{table}

   Since these values are crucial input parameters for our thermophysical
   model analysis, we repeated the search for possible shape and spin-vector
   solutions using the lightcurve inversion method developed by Kaasalainen and Torppa
   (\cite{kaasalainen01}). Our goal was to derive a set of the most likely convex
   shape models that would fit all available lightcurves. Because of the poor
   quality of some of the data, we were only able to determine the range of
   the sidereal rotation period to 7.6204--7.6510\,hours. Different shape models
   with pole orientations covering almost the entire celestial sphere (without
   any preference for pro- or retrograde solutions) fit the data equally well.
   By scanning the period-pole parameter space, we derived 77 shape models that were
   physically acceptable (rotating around the shortest axis) and for which
   the $\chi^2$ of the fit was no more than 10\% higher than the best-fit
   $\chi^2$. These models corresponded to local minima in the parameter space.

   In addition, we included both of the original Abe et al.\ (\cite{abe08}) solutions and
   added another five with the pole fixed to the two ($\lambda_{\mathrm{ecl}}$, $\beta_{\mathrm{ecl}}$)-pairs
   mentioned above and rotation periods in the range given by Abe et al.\ (\cite{abe08}).
   Only two of these models pass the $\chi^2$ + 10\% limit on the basis of the visual
   lightcurves, the other five models have higher $\chi^2$-values. Six of these
   shape models are elongated along the 'z' axis and therefore unphysical.

   For all 84 models we performed a thermophysical model analysis for
   a wide range of possible parameters (see Table~\ref{tbl:tpm_params}).

\section{Thermophysical Model Analysis}
\label{sec:tpm}

\subsection{Model and input parameters}

  The mid-IR photometric data were already described in
  Hasegawa et al.\ (\cite{hasegawa08}). The data set includes
  15 N-band Subaru observations and two dedicated {\it Akari} observations
  at 15 and 24\,$\mu$m. We binned the single-epoch {\it Spitzer} IRS data
  (Campins et al.\ \cite{campins09}) into 20 wavelength points
  (4 for band SL2, 7 for SL1, 4 for LL2, 5 for LL1;
  see Fig.~\ref{fig:irs_obs_mod}, bottom).
  The 20 wavelength points were chosen to give
  about equal weight to the two published data samples in terms
  of number of observations (17 in Hasegawa et al.\ \cite{hasegawa08}
  and 20 for the Campins et al.\ \cite{campins09} sample).
  In this way the derived object properties are better connected
  to the entire data set, and they do not just match the measurements of
  one observing epoch.
  Each observation set also has a mixture of lower and higher
  quality data: The ground-based Subaru data are of lower
  quality than the {\it Akari} data, while the quality
  of the IRS spectrum changes with wavelength. This again
  ensures that the final solutions are not biased towards
  a single measurement. All observations are listed in Table~\ref{tbl:tpm_obs}.

  \begin{table*}[h!tb]
    \begin{center}
    \caption{Summary of the avaible thermal observations. R$_{h}$ is the helio-centric
    distance and $\Delta$ the distance between object and telescope. All observations
    were taken at positive phase angles $\alpha$ (Sun-object-telescope), i.e., leading
    the Sun, with a cold terminator for a retrograde rotating body.  
             \label{tbl:tpm_obs}}
    \begin{tabular}{lllllll}
      \hline
      \hline
      \noalign{\smallskip}
      Year/Mon/Day  &  Wavelength range [$\mu$m] & R$_{h}$ [AU] & $\Delta$ [AU] & $\alpha$ [$^{\circ}$] & Telescope & Reference \\
      \noalign{\smallskip}
      \hline
      \noalign{\smallskip}
      2007/05/16 & 15.0, 24.0 & 1.414 & 0.992 & +45.6 & Akari & Hasegawa et al.\ (\cite{hasegawa08}) \\
      2007/08/28 & 8.8(3$\times$), 9.7(1$\times$), 10.5(1$\times$), 11.7(7$\times$), 12.4(3$\times$) & 1.287 & 0.306 & +22.3 & Subaru & Hasegawa et al.\ (\cite{hasegawa08}) \\
      2008/05/02 & 5.2-8.5, 7.4-14.2, 14.0-21.5, 19.5-38.0 & 1.202 & 0.416 & +52.6$^{\circ}$ & Spitzer & Campins et al.\ (\cite{campins09}) \\
      \noalign{\smallskip}
      \hline
      \noalign{\smallskip}
    \end{tabular}
    \end{center}
  \end{table*}

  All 84 possible spin-vector and shape solutions from Sect.~\ref{sec:sv} have
  been used in combination with these thermal data.

  For our analysis we are using a thermophysical model (TPM) described by
  Lagerros (\cite{lagerros96}, \cite{lagerros97}, \cite{lagerros98}) and
  M\"uller \& Lagerros (\cite{mueller98}).
  This TPM works with true illumination and observing geometries, accepts
  irregular shape models and arbitrary spin-vector solutions, works with
  roughness controlled by the r.m.s.\ of the surface slopes, considers
  heat-conduction into the surface as well as multiple scattering of both
  the solar and the thermally emitted radiation. The model has been tested
  and validated thoroughly for NEAs (e.g., M\"uller et al.\ \cite{mueller05})
  and MBAs (e.g., M\"uller \& Lagerros \cite{mueller02}).
  The TPM input parameters and applied variations are listed in Table.~\ref{tbl:tpm_params}.

  \begin{table}[h!tb]
    \begin{center}
    \caption{Summary of general TPM input parameters and applied ranges.
             \label{tbl:tpm_params}}
    \begin{tabular}{lcl}
      \hline
      \hline
      \noalign{\smallskip}
      Param.\  &  Value/Range & Remarks \\
      \noalign{\smallskip}
      \hline
      \noalign{\smallskip}
    $\Gamma$            & 0...2500                  & J\,m$^{-2}$\,s$^{-0.5}$\,K$^{-1}$, thermal inertia \\
    $\rho$              & 0.4...0.9                 & r.m.s. of the surface slopes \\
    $f$                 & 0.4...0.9                 & surface fraction covered by craters \\
    $\epsilon$          & 0.9                       & emissivity \\
    $H_{\rm{V}}$-mag.\  & 18.82\,$\pm$\,0.02\,mag       & Abe et al.\ (\cite{abe08}) \\
    G-slope             & -0.110\,$\pm$\,0.007          & Abe et al.\ (\cite{abe08}) \\
    shape               & 84 models                 & see Sect.~\ref{sec:sv} \\
    spin-axis           & 84 solutions              & see Sect.~\ref{sec:sv} \\
    P$_{\mathrm{sid}}$ [h]       & 7.6205...7.6510           & see Sect.~\ref{sec:sv} \\
     \noalign{\smallskip}
     \hline
     \noalign{\smallskip}
    \end{tabular}
    \end{center}
  \end{table}

\subsection{Solving for effective diameter, geometric albedo and thermal inertia}

  Campins et al.\ (\cite{campins09}) and Mueller (\cite{mueller07}) used
  a $\chi^2$ or reduced $\chi^2$-test to find solutions for the thermal
  inertia $\Gamma$.
  Here we follow
  a modified approach to find the most robust solutions with respect
  to thermal inertia and allowing for the full range in
  effective diameter and geometric albedo at the same time. The following
  procedure was executed for all 84 possible shape and spin-vector solutions
  separately:\\
  (1) For each value of $\Gamma$ in a wide range (see Table~\ref{tbl:tpm_params})
      we calculate the radiometric diameter and albedo solution via the
      TPM for each individually observed thermal flux (37 individual diameter and
      albedo solutions). Diameter and albedo are linked by the absolute
      magnitude $H_{\rm{V}}$ which was kept constant (the rotational amplitude is only
      about 0.1\,mag).\\
  (2) We calculate the weighted mean radiometric diameter and albedo solution
      for each given $\Gamma$ ($\bar{x} = \frac{\Sigma x_i / \sigma_i^2}{\Sigma 1 / \sigma_i^2}$,
      with diameter/albedo errors $\sigma_i$ connected to the observational errors).\\
  (3) For each individual observation we predict TPM fluxes based on the
      given $\Gamma$ and the corresponding weighted mean radiometric diameter
      and albedo from step (2).\\
  (4) The most robust solutions occur when the observations and the
      TPM predictions agree best (taking the uncertainties of the
      measurements into account in a weighted mean sense, see step 2).
      This can be expressed as $\frac{1}{N} \Sigma_{i=1}^{N} ((obs_i-mod_i)/\sigma_i)^2$,
      a modified reduced $\chi^2$ method. The most likely thermal inertia
      is found at the smallest $\chi^2$ values; the connected effective diameter
      and geometric albedo values are the ones calculated at step (2).

\begin{figure}[h!tb]
    \begin{center}
      \rotatebox{90}{\resizebox{!}{\hsize}{\includegraphics{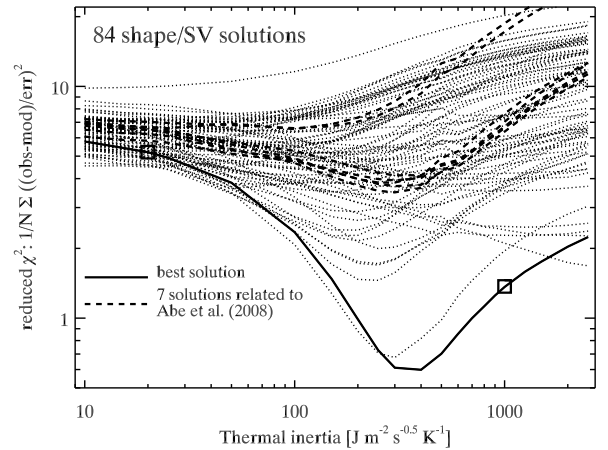}}}
      \caption{TPM $\chi^2$-optimization process to find robust solutions for diameter,
               albedo and thermal inertia simultaneously. Each line represents the 
               reduced $\chi^2$ values for an individual shape/spin-vector solution
               as a function of thermal inertia.
               The surface roughness was kept constant using the "default values"
               of $\rho$=0.7 and $f$=0.6 as proposed by M\"uller et al.\ (\cite{mueller99}).
               Model 7\_1 solutions marked with squares on the solid line
               correspond to the two cases with thermal inertias of 20 and
               1000\,J\,m$^{-2}$\,s$^{-0.5}$\,K$^{-1}$ shown in Fig.~\ref{fig:tpm_shape} (left and middle).
       \label{fig:chi2a}}
    \end{center}
  \end{figure}

  In a first round of executing this procedure we kept the surface roughness
  constant at values which were specified as "default values" for large, regolith-covered
  asteroids (M\"uller et al.\ \cite{mueller99}). The corresponding roughness
  parameter values are $\rho$=0.7, the r.m.s.\ of the surface slopes, and $f$=0.6,
  the fraction of surface covered by craters. The results are shown in Fig.~\ref{fig:chi2a}.

  The best shape and spin-vector solutions (with lowest values for the reduced
  $\chi^2$ and clear minima in Fig.~\ref{fig:chi2a}) were then the starting point
  for further tests:
    (i) Are these solutions robust against sub-sets of the thermal data?
   (ii) How does the surface roughness influence the solutions? 
  (iii) Do the solutions explain the thermal behaviour over the observed
        phase angle range (from $\sim$20$^{\circ}$ to $\sim$55$^{\circ}$)?
   (iv) Is the TPM match of equal quality at all observed wavelengths?
    (v) Are there large discrepancies at certain rotational phases?

%______________________________________________________________
\section{Results and Discussion}
\label{sec:results}

\subsection{Solution for shape and spin-vector}

  The shape and spin-vector solutions which produce
  the lowest $\chi^{2}$-values in Fig.~\ref{fig:chi2a} are
  listed in Table~\ref{tbl:sv_chi2}. The model IDs represent a full shape-model, each with
  more than 2000 surface elements and more than 1000 vertices.
  The Julian date at zero rotational phase $\gamma_0$ is in
  all cases $T_0$=2454289.0.

 \begin{table}[h!tb]
    \begin{center}
    \caption{The shape and spin-vector solutions which produce
             the lowest $\chi^{2}$-values in Fig.~\ref{fig:chi2a}.
             \label{tbl:sv_chi2}}
  \begin{tabular}{lllll}
      \hline
      \hline
      \noalign{\smallskip}
  model-ID & $\lambda_{\mathrm{ecl}}$ [$^{\circ}$] & $\beta_{\mathrm{ecl}}$ [$^{\circ}$] & P$_{\mathrm{sid}}$ [h] & $\chi^2$-min \\
      \noalign{\smallskip}
      \hline
      \noalign{\smallskip}
  7\_1\tablefootmark{a}     & 73.1 & -62.3 & 7.6323 & 0.60 \\
  8\_2\tablefootmark{a}     & 69.6 & -56.7 & 7.6325 & 0.68 \\
  14\_8              & 77.1 & -30.9 & 7.6510 & 1.46 \\
      \noalign{\smallskip}
      \hline
      \noalign{\smallskip}
  \end{tabular}
 \tablefoot{\tablefoottext{a}{Models 7\_1 and 8\_2 are in the same local
  minimum in the parameter space for the lightcurve fits.}}
 \end{center}
 \end{table}
 
  All three solutions
  are retrograde solutions and, in fact, the eight best solutions
  in Fig.~\ref{fig:chi2a} are retrograde solutions. The best prograde solutions in the
  $\chi^2$ picture are models with ID 5\_2 and ID 13\_5, both have $\chi^2$-minima at 2.3
  (almost a factor of 4 higher than the best retrograde solution 7\_1) and
  would require an extremely high thermal inertia
  ($>$ 1000\,J\,m$^{-2}$\,s$^{-0.5}$\,K$^{-1}$) to match the observations.
  Our most likely solution can be summarized as:
  $\lambda_{\mathrm{ecl}}$\,=\,73$^{\circ}$, $\beta_{\mathrm{ecl}}$\,=\,-62$^{\circ}$, P$_{\mathrm{sid}}$\,=\,7.63\,$\pm$\,0.01\,hours.
  Fig.~\ref{fig:tpm_shape} (left, middle) shows the model-ID 7\_1, as seen from
  {\it Spitzer} during the IRS-observations and for the two thermal inertias marked in
  Fig.~\ref{fig:chi2a} with boxes. The match between model ``7\_1" with observed lightcurves
  is shown in Fig.~\ref{fig:lc}.

  We also analysed the thermal data set against the Abe et al.\ (\cite{abe08}) spin-pole solutions
  discussed above. The corresponding $\chi^2$-minima are more than
  a factor five worse than our two best models above (see dashed lines in Fig.~\ref{fig:chi2a}).
  Some of these solutions are compatiple with the inertia range given by Campins
  et al.\ (\cite{campins09}) and even produced an excellent match to the IRS-spectrum.
  Nevertheless, these solutions can be excluded with very high confidence:
  (i) the match to the rest of the data set ({\it Akari} and ground-based data)
  is very poor (reflected in the high $\chi^2$-values);
  (ii) the corresponding shape models are unphysical with elongations along
  the spin axis. Such rotational states would not be stable.
  The best of these models (in terms of $\chi^2$-minima) is shown in
  Fig.~\ref{fig:tpm_shape} on the right side (with the rotation along
  the z-axis with the object's largest extension).

\subsection{Thermal inertia and surface roughness}

\begin{figure*}[h!tb]
    \begin{center}
      \rotatebox{270}{\resizebox{!}{6cm}{\includegraphics{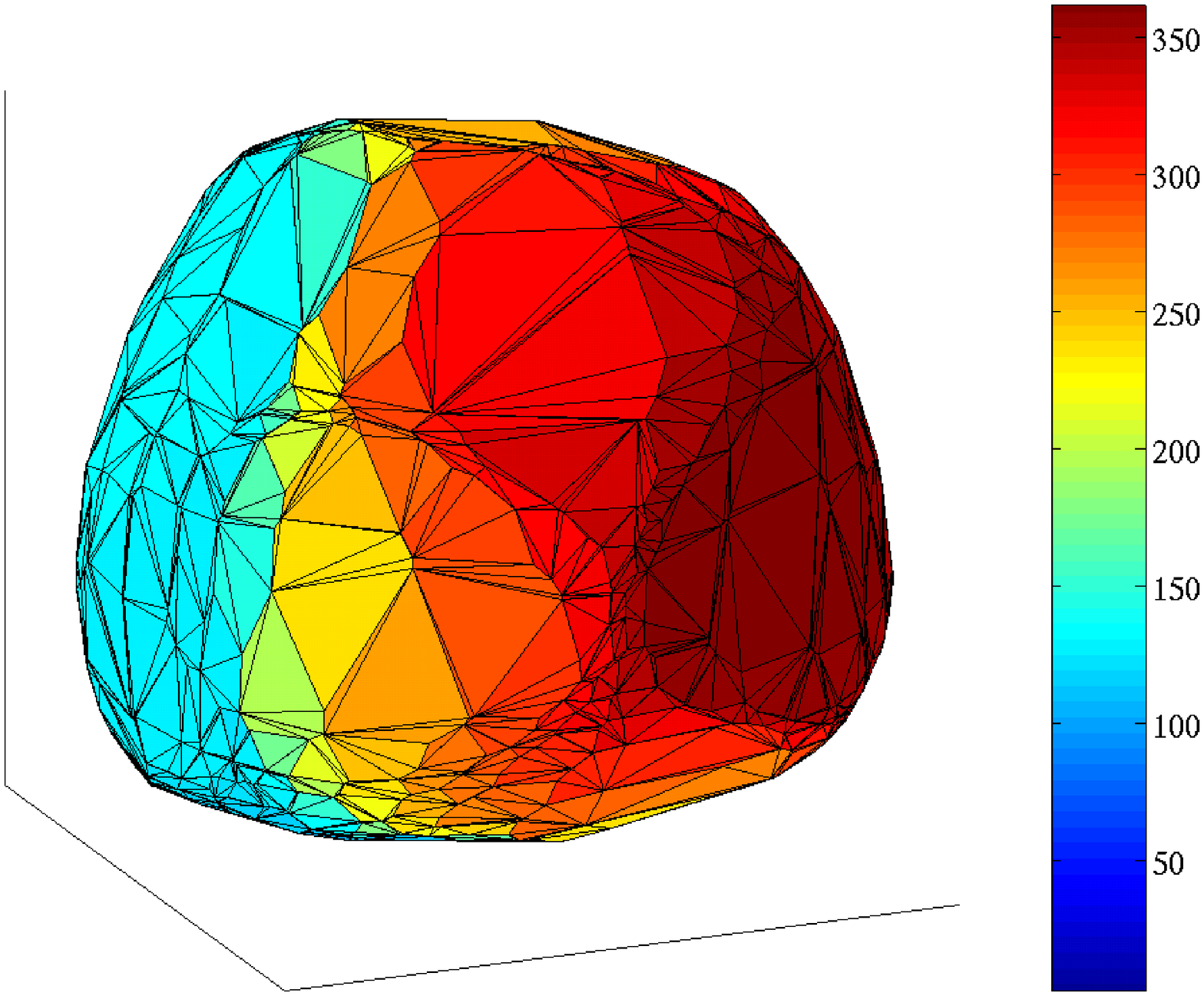}}}
      \rotatebox{270}{\resizebox{!}{6cm}{\includegraphics{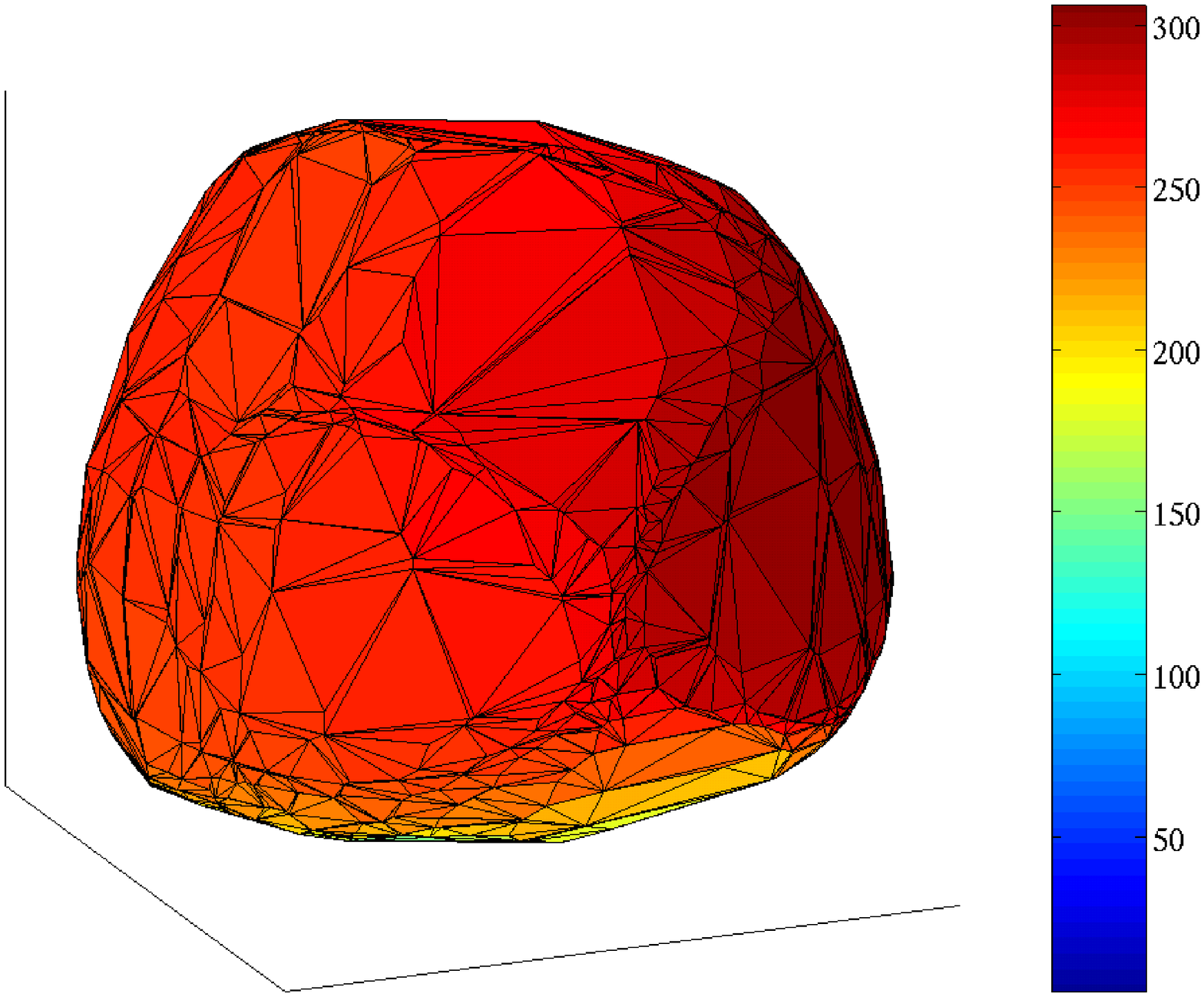}}}
      \rotatebox{270}{\resizebox{!}{6cm}{\includegraphics{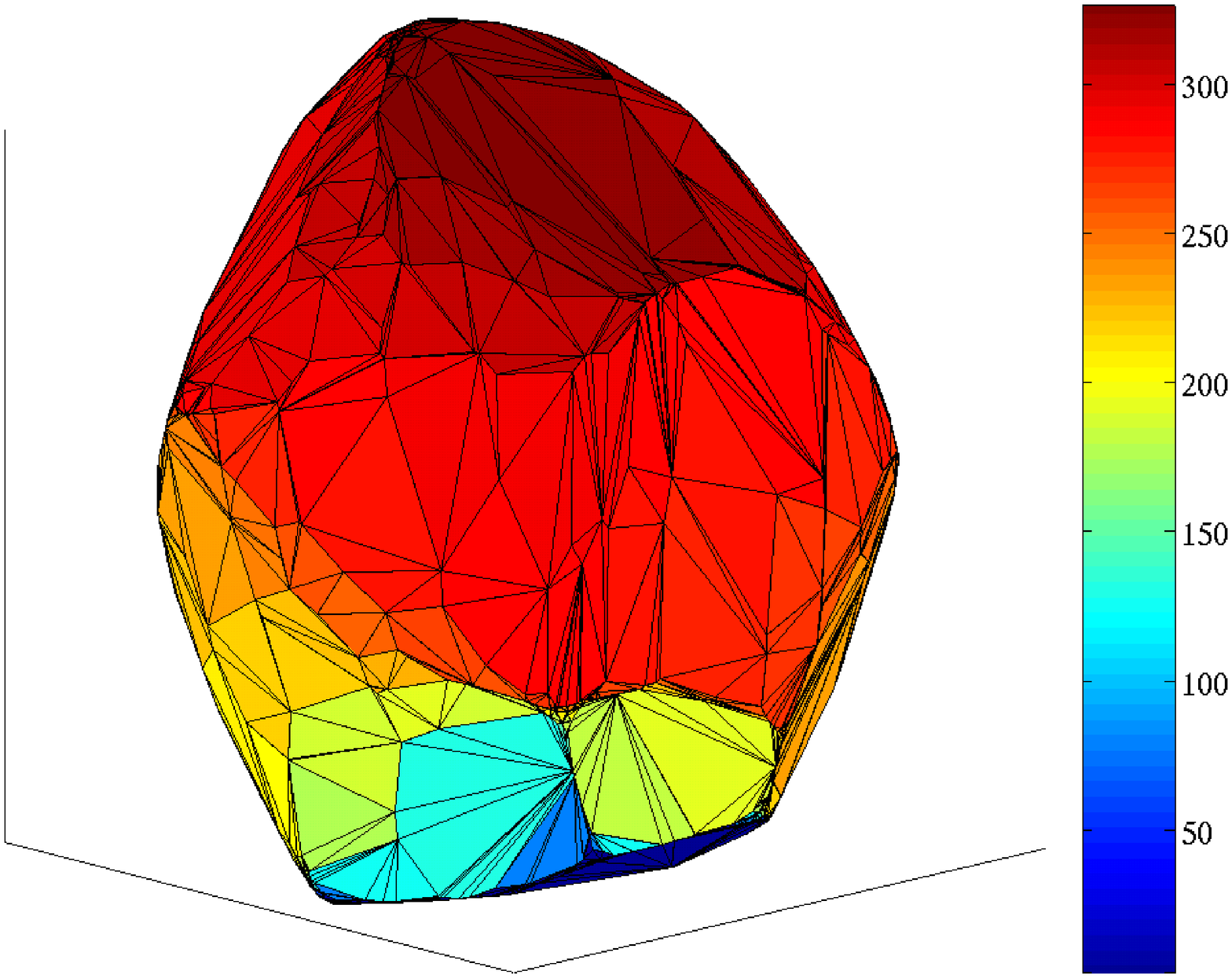}}}
      \caption{Left and middle: TPM implementation of the shape model 7\_1 at the time of the
               {\it Spitzer} IRS observations on 2008-May-02 02:01 UT, as seen 
               from {\it Spitzer} in asteroid-coordinates, i.e., the z-axis goes along
               the rotation axis. Left: low thermal inertia. Middle: high thermal
               inertia. Both solutions are marked with squares in Fig.~\ref{fig:chi2a}.
               Right: Shape model fixed on the Abe et al.\ (\cite{abe08}) spin-vector
               and tuned to match the Campins et al.\ (\cite{campins09}) findings.
               The rotation along the largest object extension (z-axis) is unphysical.
               The surface temperatures are given in Kelvin.
       \label{fig:tpm_shape}}
    \end{center}
  \end{figure*}

  The thermal inertia is clearly a key parameter when modelling the mid-IR data
  for NEAs; it strongly influences the shape of the spectral energy distribution (SED).
  This can be seen in the {\it Spitzer} IRS spectrum, especially in the Wien-part
  of the spectrum. But the thermal inertia also drives the thermal behaviour
  as a function of phase angle (e.g., M\"uller \cite{mueller02a}), the thermal phase curve.
  The temperature of the unilluminated fraction of the surface
  changes strongly with thermal inertia. In Fig.~\ref{fig:tpm_shape} the unilluminated
  fraction has just rotated out of the solar insolation. In one case (left) we assumed a low thermal
  inertia of $\sim$20\,J\,m$^{-2}$\,s$^{-0.5}$\,K$^{-1}$, which was found to be typical for
  large MBAs (M\"uller \& Lagerros \cite{mueller02}), while in the second case (middle) the
  thermal inertia is 1000\,J\,m$^{-2}$\,s$^{-0.5}$\,K$^{-1}$, close to the value found
  by M\"uller et al.\ (\cite{mueller05}) for \object{25143~Itokawa}. 

\begin{figure}[h!tb]
    \begin{center}
      \rotatebox{90}{\resizebox{!}{\hsize}{\includegraphics{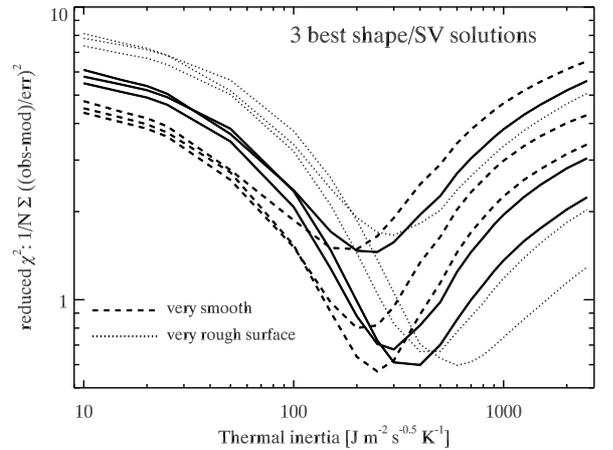}}}
      \caption{TPM $\chi^2$-optimization process for the shape and spin-axis solutions
               with the lowest $\chi^2$-values (models 7\_1, 8\_2, 14\_8).
               The solid lines are the "default roughness" values derived for MBAs.
       \label{fig:chi2b}}
    \end{center}
  \end{figure}

  The importance of the thermal inertia in the modelling is also visible
  in the $\chi^2$-solutions (Figs.~\ref{fig:chi2a} \& \ref{fig:chi2b}): the
  $\chi^2$-values change significantly when going through the whole grid
  of physically meaningful thermal inertias.
  
  But Fig.~\ref{fig:chi2b} demonstrates that thermal inertia and surface roughness
  are not easy to disentangle, at least on the basis of this data set. Both parameters influence
  the short-wavelength SED-part where the hottest surface temperatures dominate the
  SED-shape. One way of solving for both parameters would be to obtain data with a larger
  phase angle coverage and larger wavelength coverage. The roughness plays a much bigger
  role at small phase angles (beaming effect), but thermal data close to opposition
  are not available for \object{162173 (1999~JU3)}. The thermal inertia is more
  important at large phase angles and at longer wavelengths where the disk-averaged
  temperature dominates the SED shape. In general, the larger the coverage in phase
  angle and wavelength, the more accurate is the determination of diameter, albedo,
  thermal inertia and roughness.

  Fig.~\ref{fig:chi2b} also shows that similarly low $\chi^2$-values can be reached
  independent of the surface roughness. This demonstrates that roughness effects
  are still important for the interpretation of thermal data (via the thermal phase
  curves), even at these relatively large phase angles (M\"uller \cite{mueller02a}).
  But the data set does not allow us to constrain the surface roughness which broadens the 
  possible range of thermal inertias as can be seen in Fig.~\ref{fig:chi2b}.
  Based on our three best shape/spin-vector solutions and considering the uncertainties
  due to roughness, we conclude that the thermal inertia is in the range 200-600\,J\,m$^{-2}$\,s$^{-0.5}$\,K$^{-1}$.
  This range is in agreement with the lower limit of 150\,J\,m$^{-2}$\,s$^{-0.5}$\,K$^{-1}$
  given by Campins et al.\ (\cite{campins09}) but lower than their best fit
  value of 700\,$\pm$\,200\,J\,m$^{-2}$\,s$^{-0.5}$\,K$^{-1}$. Our value is about a factor
  of two lower than the one found for \object{25143~Itokawa} (M\"uller et al.\ \cite{mueller05}).
  We expect that the surface of \object{162173 (1999~JU3)} is therefore different in
  the sense that there might be fewer rocks and boulders and that the surface includes 
  millimetre sized particles (as opposed to the cm-sized gravel on Itokawa). We also find
  a rigorous lower limit to the thermal inertia of about 100\,J\,m$^{-2}$\,s$^{-0.5}$\,K$^{-1}$,
  similar to Campins et al.\ (\cite{campins09}). This limit is not compatible with a
  thick dusty regolith covering the entire surface which would result in a very
  low thermal conductivity and thermal inertia values well below 100\,J\,m$^{-2}$\,s$^{-0.5}$\,K$^{-1}$,
  which would in principle be possible for an object of that size and the relativly slow rotation rate.
  For comparison, the Moon's thick regolith gives a value below 40\,J\,m$^{-2}$\,s$^{-0.5}$\,K$^{-1}$
  (Keihm \cite{keihm84}, calculated for T = 300\,K).

\subsection{Radiometric diameter and albedo solution}

  Our best fit to all observations, as represented in Fig.~\ref{fig:chi2a} by the solid
  line, resulted in a radiometric effective diameter of 0.87\,$\pm$\,0.02\,km and
  0.070\,$\pm$\,0.003 for the visual geometric albedo. Both values are within the
  error bars of the solutions found by Hasegawa et al. (\cite{hasegawa08}) and by
  Campins et al.\ (\cite{campins09}), but now with much smaller errors.
  The uncertainties are based on the best $\chi^2$-values for model-IDs "7\_1" and
  "8\_2" and the full variation in roughness (as shown in Fig.~\ref{fig:chi2b}).
  The radiometric
  effective diameter is connected to the most likely shape model and spin-vector solution
  from above and corresponds to the size of a spherical object with equal volume.
  Using the determined possible thermal inertia range of 200-600\,J\,m$^{-2}$\,s$^{-0.5}$\,K$^{-1}$
  and uncertainties in roughness and $H_{\rm{V}}$-magnitude lead to a possible value
  for the effective diameter of 0.87\,$\pm$\,0.03\,km.
  The derived geometric albedo provides the best solution between all thermal 
  observations over the full wavelength and phase angle range and the absolute $H_{\rm{V}}$-magnitude.
  The $H_{\rm{V}}$-magnitude was given with only 0.02\,mag error. Using a more realistic
  $H_{\rm{V}}$-error of $\pm$0.1\,mag leads to a final geometric albedo value of 0.070\,$\pm$\,0.006.
  The small uncertainties reflect the importance of multi-epoch, multi-wavelength and
  large phase angle coverage for thermophysical studies of small bodies. In a similar study 
  by M\"uller et al.\ (\cite{mueller05}), also based on a large thermal data set and a shape
  model from lightcurve inversion techniques, the derived effective diameter agreed within
  2\% of the true, in-situ diameter (M\"uller et al.\ \cite{mueller05}; Fujiwara et
  al.~\cite{fujiwara06}). The quoted uncertainties above are formal errors from the $\chi^2$
  optimization, including the possible range in thermal inertia, roughness and $H_{\rm{V}}$.

\begin{figure}[h!tb]
    \begin{center}
      \rotatebox{90}{\resizebox{!}{\hsize}{\includegraphics{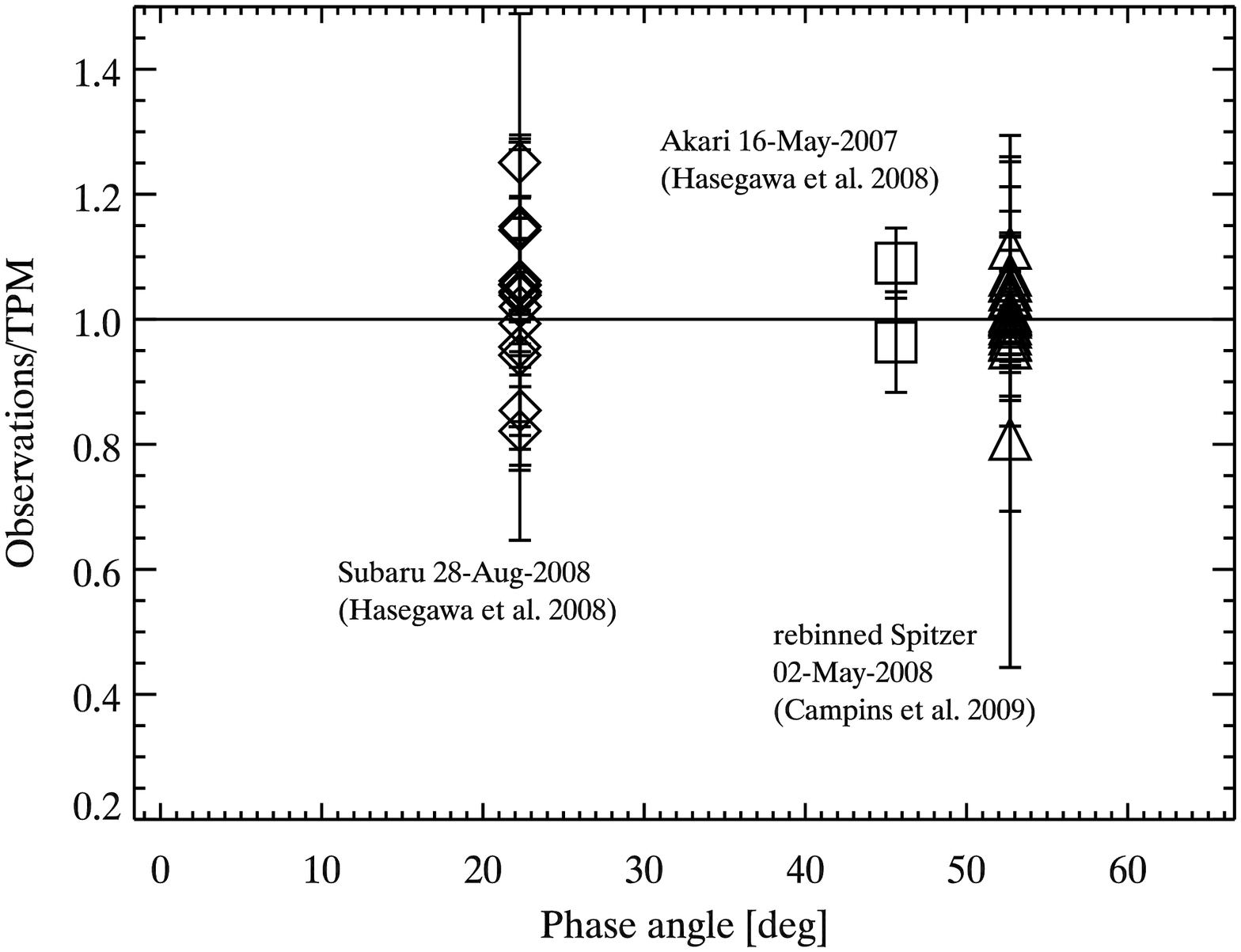}}}
      \rotatebox{90}{\resizebox{!}{\hsize}{\includegraphics{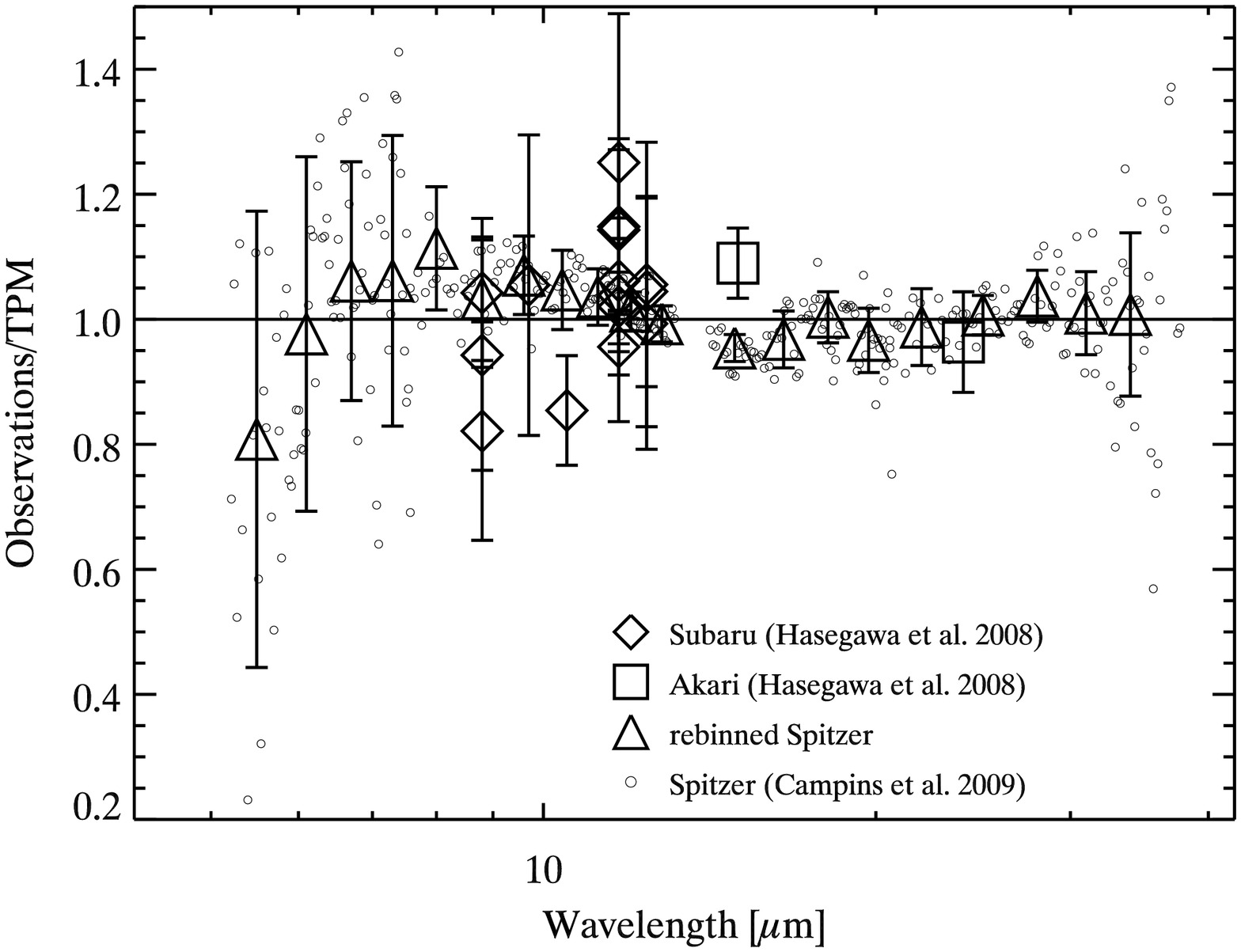}}}
      \caption{All observations divided by the corresponding TPM predictions based on our
               optimized radiometric solution. Top: as a function of phase angle. Bottom: as a
               function of wavelength.
               The full set of {\it Spitzer} IRS data are shown with little circles
               (Campins et al.\ \cite{campins09}), the triangles are the re-binned
               data, the {\it Akari} IRC data are represented by squares; the
               Subaru-COMICS observations by diamond symbols.
       \label{fig:irs_obs_mod}}
    \end{center}
  \end{figure}

  The remaining discrepancy between the {\it Spitzer} and the {\it Akari}-flux at 15\,$\mu$m
  (seen in Fig.~\ref{fig:irs_obs_mod} bottom) might be related to an additional error
  introduced by the flux scaling done by Campins et al.\ (\cite{campins09}) to match the short-wavelength part
  of the spectrum ($<$14\,$\mu$m) to the long-wavelength part of the spectrum ($>$14\,$\mu$m).
  The mismatch is caused mainly by the placement of the object within the IRS slit and
  reflected in the specified 10\% systematic absolute calibration uncertainty.

  Although the Fig.~\ref{fig:chi2b} solutions and the model match in
  Fig.~\ref{fig:irs_obs_mod} look very convincing,
  there are some uncertainties remaining: The spin-vector and
  shape solution from lightcurve inversion techniques is not
  very robust; more lightcurve observations are needed to
  improve the quality. The $\chi^2$-test works best if the thermal
  observations cover a wide range of wavelengths, phase angles (before and after
  opposition) and rotational phases. But all thermal observations have
  been taken at pre-opposition (positive) phase angles (leading
  the Sun) where the unilluminated part of the surface 
  visible to the observer is either warm (prograde rotation)
  or cold (retrograde rotation); observations after opposition are not yet available.
  A combined before and after opposition data set would constrain the sense of
  rotation and the thermal properties much better, observations close to opposition
  would determine the surface roughness a bit better, hence constrain the
  thermal inertia further.

  It is also important to note here that higher thermal inertias
  ($>$600\,J\,m$^{-2}$\,s$^{-0.5}$\,K$^{-1}$, Fig.~\ref{fig:tpm_shape} middle) would make a slightly
  better fit to the {\it Spitzer} IRS spectrum (improvement mainly at the
  shortest wavelengths in Fig.~\ref{fig:irs_obs_mod} bottom),
  but would cause a significant dependency
  in the diameter and albedo solutions with phase angle. The measurements
  taken at around 20$^{\circ}$ phase angle (Subaru) would then have fluxes 
  that are about 40\% higher than the corresponding model predictions
  (i.e., values $>$1.4 in Fig.~\ref{fig:irs_obs_mod}, top).
  The {\it Akari} fluxes would still be $\sim$ 20\% higher than the model
  predictions. Our observational data set covering about 30$^{\circ}$ in
  phase angle constrains the possible thermal inertia to values
  below 700\,J\,m$^{-2}$\,s$^{-0.5}$\,K$^{-1}$.

\section{Conclusions}

  The radiometric analysis provides the following results:
  (i)   a strong indication of a retrograde sense of rotation;
  (ii)  a spin-vector with  $\lambda_{\mathrm{ecl}}$\,=\,73$^{\circ}$,
        $\beta_{\mathrm{ecl}}$\,=\,-62$^{\circ}$, P$_{\mathrm{sid}}$\,=\,7.63\,$\pm$\,0.01\,h and
        $\gamma_0$\,=\,0 at T$_0$\,=\,2454289.0, derived for the first time based on a
        combined analysis of visual lightcurve data {\bf and} thermal observations;
  (iii) a shape model (here labelled with 7\_1) as shown in Fig.~\ref{fig:tpm_shape}
        (left \& middle);
  (iv)  a thermal inertia in the range 200 to 600\,J\,m$^{-2}$\,s$^{-0.5}$\,K$^{-1}$;
  (v)   a radiometric effective diameter (of an equal volume sphere)
        of D$_{\mathrm{eff}}$ = 0.87\,$\pm$\,0.03\,km;
  (vi)  a radiometric geometric albedo of p$_{V}$\,=\,0.070\,$\pm$\,0.006;
  (vii) a lower thermal inertia than for Itokawa, suggesting the presence of 
        smaller particles, $<$ cm-sized, in the regolith, though likely not fine dust;
  (viii) very good agreement in the radiometric solutions between
        the {\it Spitzer}, the {\it Akari} and the {\it Subaru} observations;
  (ix) an excellent match of the flux changes with phase angle
       (the phase angle range covered here is from $\sim$20$^{\circ}$ to $\sim$55$^{\circ}$).

  The example of \object{162173 (1999~JU3)} shows that a combination of
  visual lightcurves (reflected sunlight) and mid-/far-IR photometry or
  photo-spectroscopy (thermal emission) can improve the quality of shape and spin-vector
  solutions significantly.

\begin{acknowledgements}
      J.\ D.\ received grants from the Czech Science Foundation (GACR P209/10/0537) and the
      Research Program MSM0021620860 of the Ministry of education.
      S.\ H.\ was supported by the Space Plasma Laboratory, ISAS, JAXA.
      We are also grateful to Professor N.\ Kawai and the gamma-ray
      bursts project members for furnishing their optical camera at
      Ishigakijima Astronomical Observatory. Development of the optical
      CCD camera at Ishigakijima Astronomical Observatory was supported
      by the Ministry of Education, Science, Sports and Culture,
      Grant-in-Aid for Creative Scientific Research.
      T.\ K.\ thanks to the JSPS Research Fellowships for Research Abroad
      for their financial support.
      This work was supported in part by the NASA Planetary Astronomy
      Program and performed in part at the Jet Propulsion Laboratory. We
      thank the Steward Observatory of the University of Arizona for its
      allocation of telescope time.
      We also thank the referee Dr.\ J.\ Emery for very helpful comments.
\end{acknowledgements}


\begin{thebibliography}{}

\bibitem[2008]{abe08}
          Abe, M., Kawakami, K., Hasegawa, S.\ et al.\ 2008,
          COSPAR Scientific Assembly, B04-0061-08
\bibitem[2004]{binzel04}
          Binzel, R.\ P., Perozzi, E., Rivkin, A.\ S., Rossi, A.,
          Harris, A.\ W., Bus, S.\ J., Valsecchi, G.\ B., Slivan, S.\ M.\ 2004,
          Meteorit.\ Planet.\ Sci., 39, 351

\bibitem[2009]{campins09}
         Campins, H., Emery, J.P., Kelley, et al.\ 2009,
%        Spitzer observations of spacecraft target 162173 (1999~JU3)
         A\&A 503, L17-L20
	 
\bibitem[2006]{fujiwara06}
         Fujiwara, A., Kawaguchi, J., Uesugi, K.\ et al.\ 2006, \textit{Science} 312, 1330

\bibitem[2008]{hasegawa08}
         Hasegawa, S., M\"uller, T.\ G., Kawakami, K., Kasuga, T.,
         Wada, T., Ita, Y., Takato, N., Terada, H., Fujiyoshi, T.,
         Abe, M.\ 2008,
%        Albedo, Size, and Surface Characteristics of Hayabusa-2 Sample-Return
%        Target 162173 1999~JU3 from AKARI and Subaru Observations,
         PASJ 60, 399

\bibitem[2001]{kaasalainen01}
         Kaasalainen, M.\, Torppa, J.\ 2001,
         Icarus 153, 24

\bibitem[1984]{keihm84}
         Keihm, S.\ J.\ 1984,
         Icarus 60, 568

\bibitem[1996]{lagerros96}
         Lagerros, J.\ S.\ V.\ 1996, A\&A 310, 1011
\bibitem[1997]{lagerros97}
         Lagerros, J.\ S.\ V.\ 1997, A\&A 325, 1226
\bibitem[1998]{lagerros98}
         Lagerros, J.\ S.\ V.\ 1998, A\&A 332, 1123

\bibitem[1986]{magnusson86}
        Magnusson. P.\ 1986, Icarus 68, 1-39
% Distribution of Spin Axes and Senses of Rotation for 20 Large Asteroids,

\bibitem[1998]{mueller98}
          M\"uller, T.\ G.\ \& Lagerros, J.\ S.\ V.\ 1998,
          A\&A 338, 340

\bibitem[1999]{mueller99}
         M\"uller, T.\ G., Lagerros, J.\ S.\ V., Burgdorf, M.\ et al.\ 1999,
         ESA SP-427, in The Universe as Seen by ISO,
         P.\ Cox \& M.\ F.\ Kessler (Eds.), 141

\bibitem[2002]{mueller02}
          M\"uller, T.\ G.\ \& Lagerros, J.\ S.\ V.\ 2002,
          A\&A 381, 324

\bibitem[2002]{mueller02a}
         M\"uller, T.\ G.\ 2002,
         M\&PS 37, 1919

\bibitem[2005]{mueller05}
         M\"uller, T.\ G., Sekiguchi, T., Kaasalainen M., Abe M., Hasegawa S.\ 2005,
         A\&A 443, 347 (M05)

\bibitem[2007]{mueller07}
         Mueller, M.\ 2007,
         % Surface Properties of Asteroids from Mid-Infrared Observations and Thermophysical Modeling,
         DLR, Thesis FU Berlin
         (http://www.diss.fu-berlin.de/diss/receive/FUDISS\_thesis\_000000002596)

\end{thebibliography}
\end{document}